\RequirePackage{lineno}
\documentclass[10pt,numberedappendix]{emulateapj}

\usepackage[normalem]{ulem}
\usepackage{booktabs}
\usepackage[utf8]{inputenc}
\usepackage{graphicx}%
\usepackage{dcolumn}%
\usepackage{bm}%
\usepackage{amsmath}
\usepackage{xcolor}
\usepackage{xspace}
\usepackage{hyperref}
\usepackage{url}
\usepackage[utf8]{inputenc}

\newif\ifthetajn \thetajnfalse

\begin{document}

\title{Searching for exotic cores with binary neutron star inspirals}

\author{Hsin-Yu~Chen}
\email{hsinyuchen@fas.harvard.edu}

\author{Paul~M.~Chesler}
\email{pchesler@g.harvard.edu}

\author{Abraham~Loeb}
\email{aloeb@cfa.harvard.edu}

\affiliation{Black Hole Initiative, Harvard University, Cambridge, Massachusetts 02138, USA}

\begin{abstract}
We study the feasibility of detecting exotic cores in merging
neutron stars with ground-based gravitational-wave detectors. We focus on models with a sharp nuclear/exotic matter interface, and assume a uniform distribution of neutron stars in the mass range $[1,2] M_\odot$. We find that the existence of exotic cores can be confirmed at the 70\% confidence level with as few as several tens of detections.  Likewise, with such a sample, we find that some models of exotic cores can be excluded {with high confidence}.
\end{abstract}

\maketitle

\section{Introduction}The discovery of binary neutron star mergers (BNSs) by 
Advanced LIGO-Virgo~\citep{2017PhRvL.119p1101A} has ushered an era of multimessenger astronomy~\citep{2017ApJ...848L..12A}. 
BNSs also provide a unique laboratory to study the equation of state (EoS) of Quantum Chromodynamics (QCD) matter, which at densities beyond the nuclear saturation density is largely unknown.  Already, the single event GW170817 indicates a relatively soft EoS \citep{2018PhRvL.121p1101A,2019PhRvX...9a1001A}. Advanced LIGO-Virgo is expected to observe tens to hundreds of BNSs in the next few years~\citep{2018LRR....21....3A}, substantially improving constraints on the EoS of dense QCD matter over time (see \textit{e.g.} \citet{Annala:2017llu,Most:2018hfd,Tews:2019cap,Forbes:2019xaz,Carson:2019xxz}).  

At densities beyond the nuclear saturation density, microscopic calculations of the EoS are increasingly uncertain and dense QCD matter is likely strongly coupled. 
Because of this, the nature of matter in the cores of NSs is largely inaccessible with current theoretical tools.  An exciting possibility is the existence of exotic matter (EM) at high densities, including hyperonic and quark matter.  This can result in 
hybrid neutron stars (HNS), with an EM core surrounded by a mantle of nuclear matter (see \textit{e.g.} \cite{1971SvA....15..347S,1983A&A...126..121S,Lindblom:1998dp,Yagi:2016bkt,Alford:2001zr,Alford:2015gna,Alford:2013aca,Annala:2019puf,Alford:2019oge,Haensel:2016pjp,Fortin:2017dsj,2019PhRvD..99j3009M}).  EM can also be 
produced during post-merger dynamics \citep{Bauswein:2018bma,Most:2018eaw}.  A tantalizing question  
arises as to whether the effects of EM be observed with gravitational wave (GW) detectors.  If so, what are the observable signatures of EM? How many mergers  must one observe before
the presence of EM can be confidently established?  These are the questions we seek to address.  
 
We focus on EM scenarios most easily observable with 
current GW detectors.  As no obvious sign of 
post-merger dynamics were recorded in GW170817, likely due to post-merger dynamics 
being outside of LIGO's frequency bandwidth, we choose to study the effects of HNSs 
during the inspiral phase of NS mergers.  Additionally, we focus on HNSs with a sharp transition
between nuclear and EM, meaning those without a mixed phase.

\begin{figure}
	\begin{center}
		\includegraphics[width=0.45\textwidth]{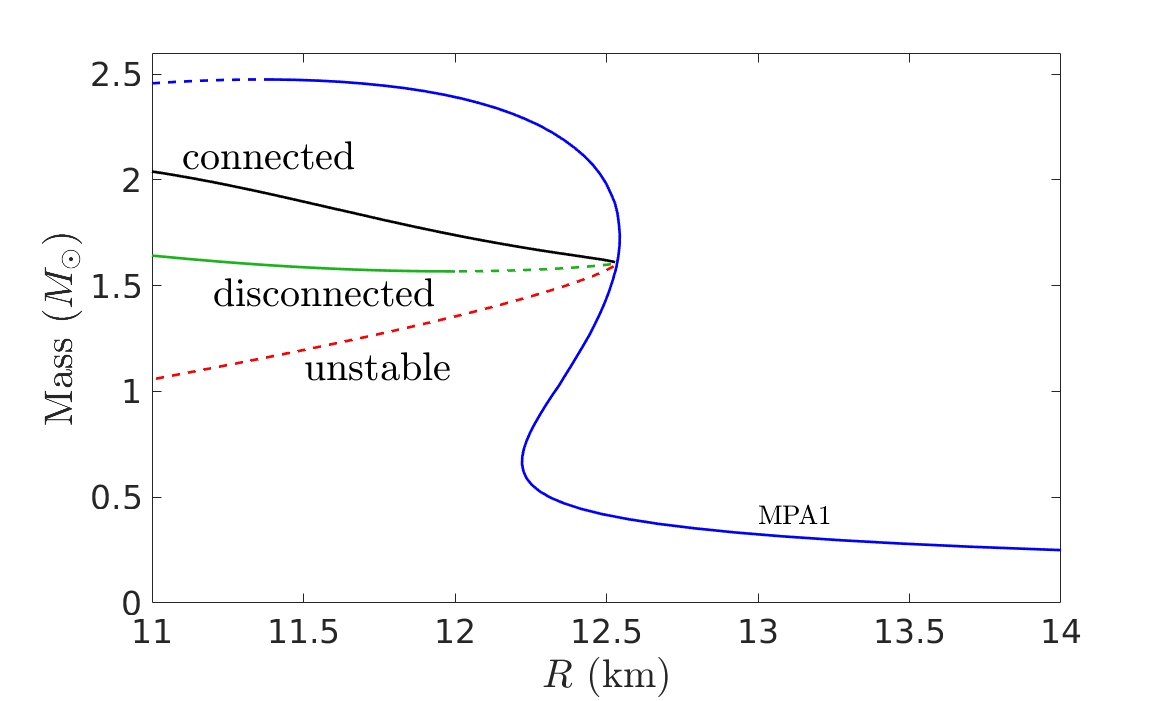}
		\caption{ An example of MR curves with and without EM.  The critical mass for the hybrid branch is $m_c = 1.6 M_\odot$, beyond which EM cores exists.
	    Regions with positive slope are unstable and shown as dashed lines.}
		\label{fig:3cases}
	\end{center}
\end{figure}

The presence of a EM core 
can have dramatic effects on mass-radius (MR) 
curves, with the nature of the modification depending on the discontinuity in the energy density (\textit{i.e.} latent heat) and the speed of sound \citep{Alford:2013aca,Alford:2015gna,Steiner:2017vmg,Benic:2014jia}.  For illustrative purposes, Fig.~\ref{fig:3cases} shows the MR curve for the MPA1 nuclear EoS of ~\cite{Muther:1987xaa} together 
with hybrid MR curves generated via a constant sound speed EM EoS.
If the latent heat is large (or the sound speed is small), HNSs are unstable to collapse and the stable branch of the MR curve terminates at some mass $m_c$, which is the mass where EM is first nucleated \citep{1983A&A...126..121S,Lindblom:1998dp}. For smaller latent heats (or larger sound speeds), the resulting HNSs can have both connected and disconnected MR curves. Crucial to our analysis is the fact that the slope of hybrid branch of the MR curve need not be the same as that of the  nuclear branch.

We focus on mergers with component masses in the $[1,2] M_\odot$ range.
There are two reasons for doing this. First, the heaviest observed neutron stars (NSs) have
masses $2.01 \pm 0.04 M_\odot$ \citep{Antoniadis:2013pzd}
and $2.17 \pm 0.1 M_\odot$ \citep{Cromartie:2019kug}.  
Because $2 M_\odot$ NSs are evidently stable, astrophysical processes such as supernova and NS mergers presumably cannot produce black holes lighter than $2 M_\odot$.  Likewise, $ \mathcal O(M_\odot)$ primordial black holes are not expected to be abundant (see for example \cite{Carr:2018poi}). 
It is therefore reasonable to surmise that all mergers with components
in the $[1,2] M_\odot$ range are BNSs. If the BNS detections are contaminated by neutron star-black hole mergers, significant biases in measured radii can be introduced \citep{2019arXiv190311197C}.  
Second, in the $[1,2] M_\odot$ range
the vast majority of viable nuclear EoS 
yield roughly constant MR curves (see for example
Fig.~10 of ~\cite{Ozel:2016oaf}).  
This should be contrasted with the kink at critical mass $m_c$ shown in Fig.~\ref{fig:3cases}.  Restricting our attention to the $[1,2] M_\odot$ range therefore allows us to use abrupt changes  in MR curves -- kinks -- as an identifier for exotic cores, with the location of the kink corresponding to the critical mass $m_c$.  
As we shall see below, kinks in MR curves readily manifests themselves in GW observables, and can be seen with a large enough collection of  events.

\section{Simulations} We construct an ensemble of simulated BNSs.
Input data consists of the NS masses $m_1$ and  $m_2$ and the NS MR curve.  
As the mass distribution of NSs outside the Milky Way is unknown, we choose $m_1$ and $m_2$ uniformly distributed in $[1,2] M_\odot$, with $m_1 \geq m_2$. Additional GW observables we employ are the chirp mass
$\mathcal M \equiv {(m_1 m_2)^{3/5}}/{(m_1 + m_2)^{1/5}}$,
and the mass weighted tidal deformablity 
\begin{equation}
\label{eq:tildeLambda}
\widetilde \Lambda \equiv \frac{16}{13} 
\frac{ ( m_1 + 12 m_2)m_1^4 \Lambda_1 + ( m_2 + 12 m_1)m_2^4 \Lambda_2}{(m_1 + m_2)^5},
\end{equation}
where $\Lambda_1$ and $\Lambda_2$ are the tidal deformablities of the individual NSs. $\widetilde \Lambda$ is determined from the MR curve
and the masses $m_1$ and $m_2$.  This follows from the universal relationship between each NS's compactness 
$C \equiv m/R$, and tidal
deformablity $\Lambda$, which reads \citep{Maselli:2013mva,Yagi:2016bkt}
\begin{equation}
\label{eq:CLrelation}
C = a_0 + a_1 \log \Lambda + a_2 \log^2 \Lambda,
\end{equation}
where $a_0 = 0.3617$, $a_1 = -0.03548$ and $a_2 = 6.194 \times 10^{-4}$. Eq.~(\ref{eq:CLrelation}) holds at the 7\% level or better for both purely nuclear NSs and HNSs \citep{Carson:2019rjx}.  
Eq.~(\ref{eq:CLrelation}) can be inverted to find $\Lambda(C)$, which
upon substituting into Eq.~\ref{eq:tildeLambda} yields $\widetilde \Lambda$ 
in terms of $m_1$ and $m_2$ and the NS radii $R_1$ and $R_2$. 
Since Eq.~\ref{eq:CLrelation} is insensitive to the underlying EoS, we
can directly apply it to the MR curves employed in the following
two paragraphs to obtain $\widetilde \Lambda$ for a given BNS with mass
$(m_1,m_2)$.

\begin{table}
	\centering
	\begin{tabular}{c c  c  c c c}
		& Model  & $p_0 ({\rm km})$ \  & $p_1 (M_\odot)$ \ & $p_2 ({\rm km}/M_\odot)$ \ & $p_3({\rm km}/M_\odot) $   \\
		\hline
		& A & 12.5 &2.0 &0.3 &0.0  \\ 
		& B & 12.5 &1.5 &0.3 &-0.7 \\ 
		& C & 12.5 &1.8 &0.8 &-0.3   \\ 
		& D & 12.5 &1.8 &0.6 &-5.0   \\ 
	    & E & 13.32 &1.6 &0.9 &-1.7   \\ 
		& F & 12.55 &1.75 &0.3 &-0.4   \\ 
		& G & 12.1 &1.5 &0.2 &-0.35   \\ 	
		& H & 11.8 &1.4 &-0.4 &-1.4   
	\end{tabular}
	\addtolength{\tabcolsep}{-10pt}
	\caption{Parameters chosen for our nuclear MR curves.}\label{eos}
\end{table}

\begin{figure}
	\centering
	\includegraphics[width=1.0\columnwidth]{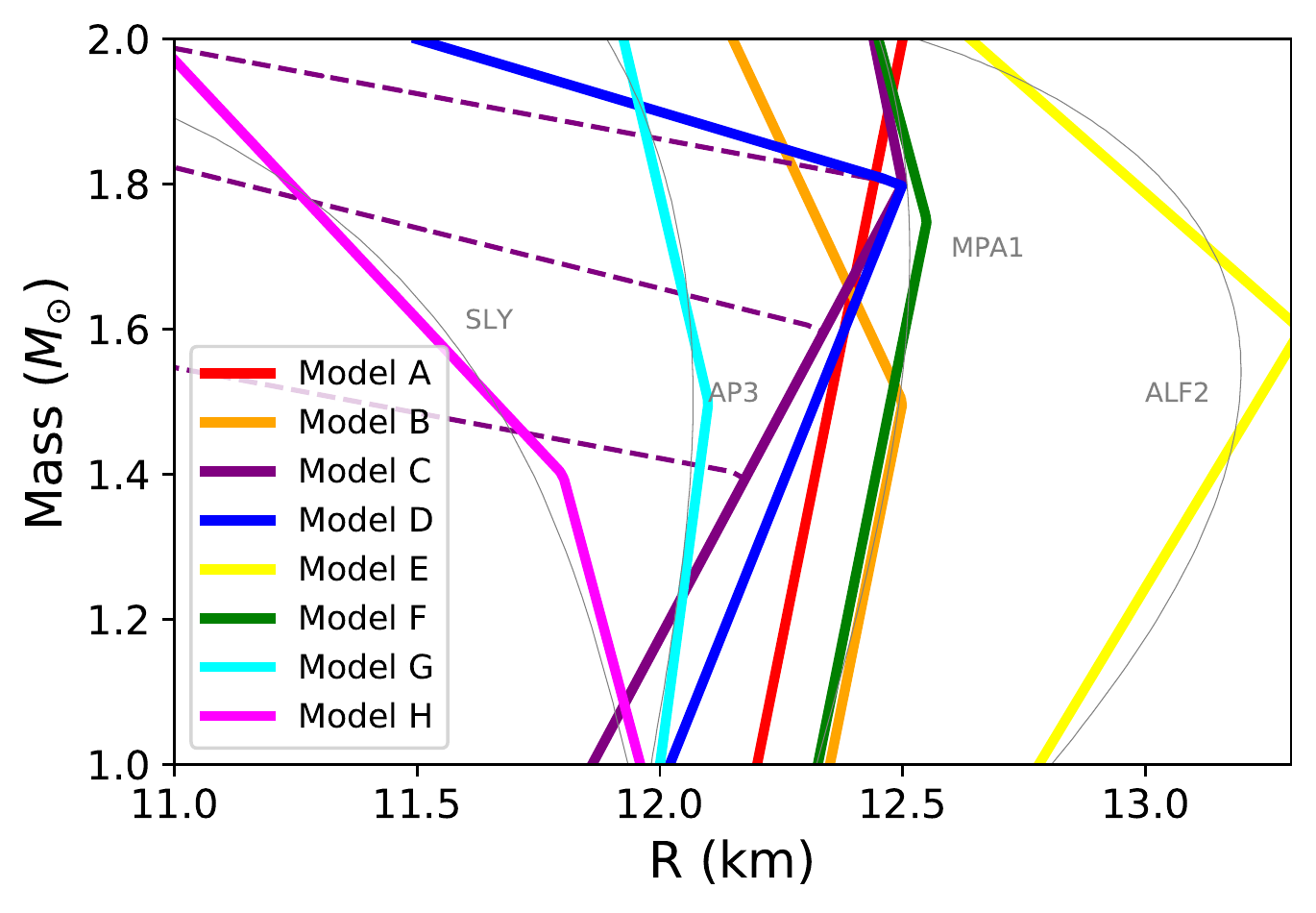}
	\caption{\label{fig:mr}
		Eight linear piecewise models for $R_{\rm nuclear}(m)$.  For comparison, the MR curves generated with several candidate nuclear EoS are also shown.  The purple dashed lines correspond to several hybrid branches of the MR curves.
	}
\end{figure}

Our analysis below is unable to resolve detailed structure in MR curves.  For simplicity we therefore employ piecewise 
linear MR curves.  For nuclear MR curves we use $R_{\rm nuclear}(m) = \mathcal R(m|p_0,p_1,p_2,p_3)$ where
\begin{equation}\label{eq:pw}
\mathcal R(m|p_0,p_1,p_2,p_3) \equiv 
\begin{cases}
p_2 (m-p_1) +p_0, &  m<p_1, \\
p_3 (m-p_1) +p_0, &  m \geq p_1.
\end{cases}
\end{equation}
We employ eight sets of parameters $p_i$ (models A-H) listed in Table~\ref{eos}, with the associated MR
curves  plotted in Fig.~\ref{fig:mr}.  Note our analysis below is insensitive to 
overall radial shifts (\textit{i.e.} the parameter $p_0$).
For comparison, also shown in the figure are several MR curves obtained from ~\cite{Muther:1987xaa,Douchin:2001sv,Akmal:1998cf,Alford:2004pf}.  Note models 
E-H approximate these MR curves.  

We focus on HNS with connected MR curves with
\begin{equation}
R_{\rm hybrid}(m) \equiv 
\begin{cases}
R_{\rm nuclear}(m), &  m<m_{c}, \\
\alpha (m - m_c) + R_{\rm nuclear}(m_c), &  m \geq m_c,
\end{cases}
\end{equation} 
with critical mass $m_c$ and hybrid branch slope $\alpha$
\begin{subequations}
	\begin{align}
	m_c &= \{1.2,1.4,1.6,1.8\} M_{\odot}, \\
	-\alpha &= \{0.5,1,2,4,6,8,10\}{\rm km}/{M_\odot}.
	\end{align}
\end{subequations}
We use the same parameters $p_i$ listed in Table~\ref{eos} and refer to hybrid MR curves as ``hybrid models A-H with hybrid branch
slope $\alpha$ and critical mass $m_c$"~\footnote{Some of the parameter combinations could lead to hybrid models 
with radii less than the Schwarzschild radii. We removed these models in our study.}.  Also shown in Fig.~\ref{fig:mr} are several hybrid branches for model C.

The signal-to-noise ratio (SNR) of the simulated detections follows a power-law distribution valid for nearby events with redshift less than $\sim 0.1$ (see ~\cite{2011CQGra..28l5023S,2014arXiv1409.0522C} for details). 
For each merger we use the SNR to add Gaussian noise to the \textit{inferred} radius, Eq.~(\ref{eq:radiusmeasurement}) below, and masses $m_1$  such that a GW170817-like event has 1-$\sigma$ radius and mass uncertainties of 0.75 km and $0.1 M_{\odot}$, respectively \citep{2018PhRvL.121p1101A,2019PhRvX...9a1001A}.
The chirp mass is assumed to be measured with negligible uncertainty.

\section{Data analysis and results}
\begin{figure}
	\begin{center}
		\includegraphics[trim= 0 0 0 0,clip,scale=0.45]{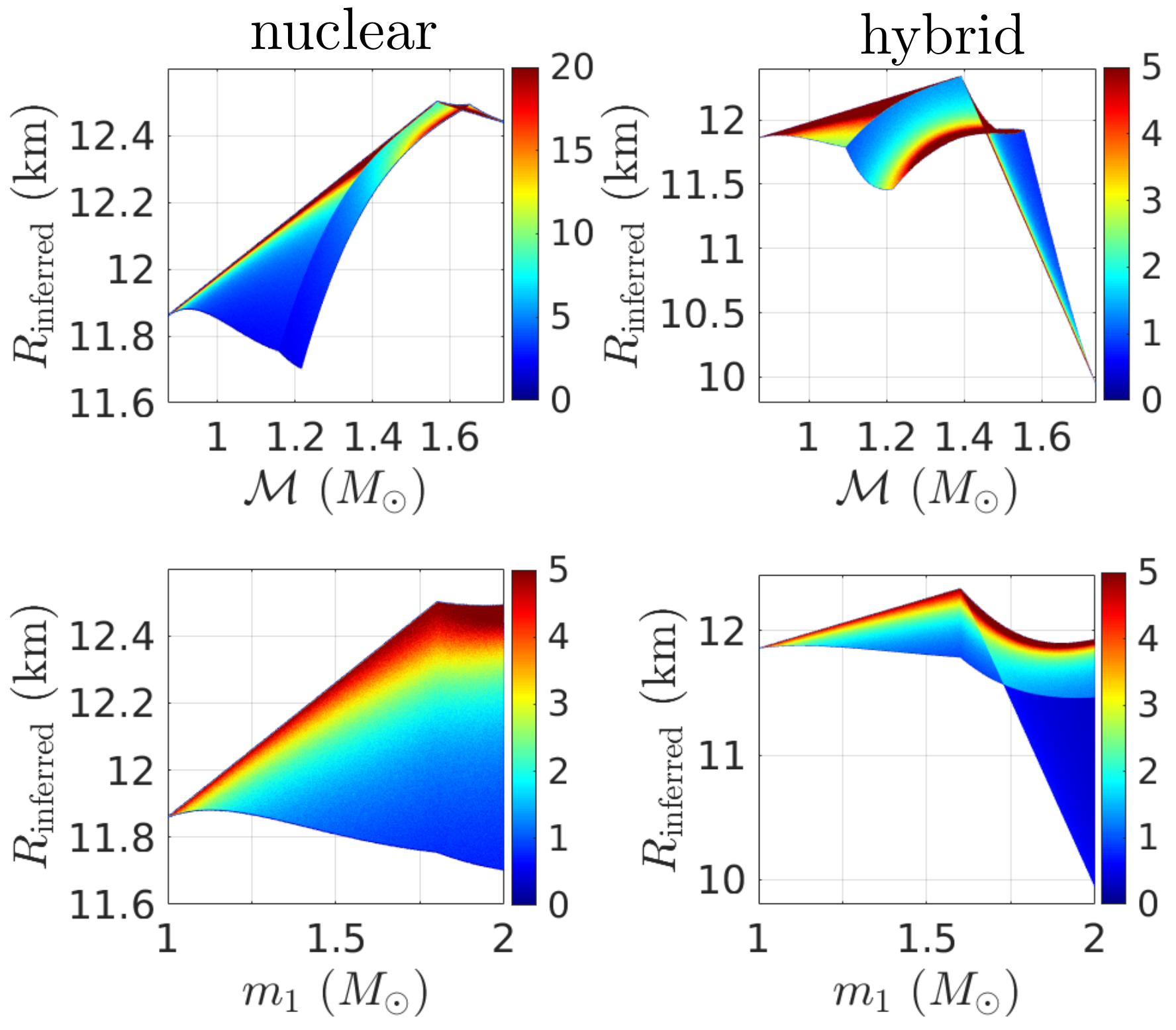}
		\caption{The density of events in the $\mathcal M{-}R_{\rm inferred}$ (top)and $m_1{-}R_{\rm inferred}$ (bottom) planes for nuclear model C (left) and a hybrid model (right) with hybrid branch slope $\alpha = -6$ and critical mass $m_c = 1.6 M_\odot$. Outside the shaded regions the densities vanish. 
        Note the presence of kinks in the hybrid allowed regions, which occur at $\mathcal M = m_c/2^{1/5}$ and $m_1 = m_c$.} 
		\label{fig:MChirpR}
	\end{center}
\end{figure}

A useful observable to analyze data is the 
\textit{inferred} radius,
\begin{equation}
\label{eq:radiusmeasurement}
R_{\rm inferred}(\mathcal M,\widetilde \Lambda) \equiv {2^{1/5} \mathcal M}/{C(\widetilde \Lambda)},
\end{equation}
where $C(\Lambda)$ is given by Eq.~(\ref{eq:CLrelation}). 
Note that in the special case $m_1 = m_2$, 
Eqs.~(\ref{eq:tildeLambda}) and (\ref{eq:CLrelation}) imply $R_1 = R_2 = R_{\rm inferred}$.
For our analysis, the utility of $R_{\rm inferred}$ lies in the fact 
that it is sensitive to kinks in MR curves.
To illustrate this, in Fig.~\ref{fig:MChirpR} we show 
the density of events in the $\mathcal M{-}R_{\rm inferred}$ (top)
and $m_1{-}R_{\rm inferred}$ (bottom) planes
for nuclear model C (left) and hybrid model C with critical mass $m_c = 1.6 M_\odot$ and hybrid branch slope $\alpha = -6 \ {\rm km}/M_\odot$.  The density of events vanishes outside the shaded regions 
\footnote
{
	Note the densities in some regions go off the color scale.
}.
As is evident from the figure,
HNSs produce pronounced kinks in allowed regions, with negative slopes at larger $\mathcal M$ and $m_1$. In contrast, the allowed regions generated by nuclear model C show no prominent kinks or pronounced downwards trends as $\mathcal M$ or $m_1$ increase.

\begin{figure}
	\centering
	\includegraphics[width=1.0\columnwidth]{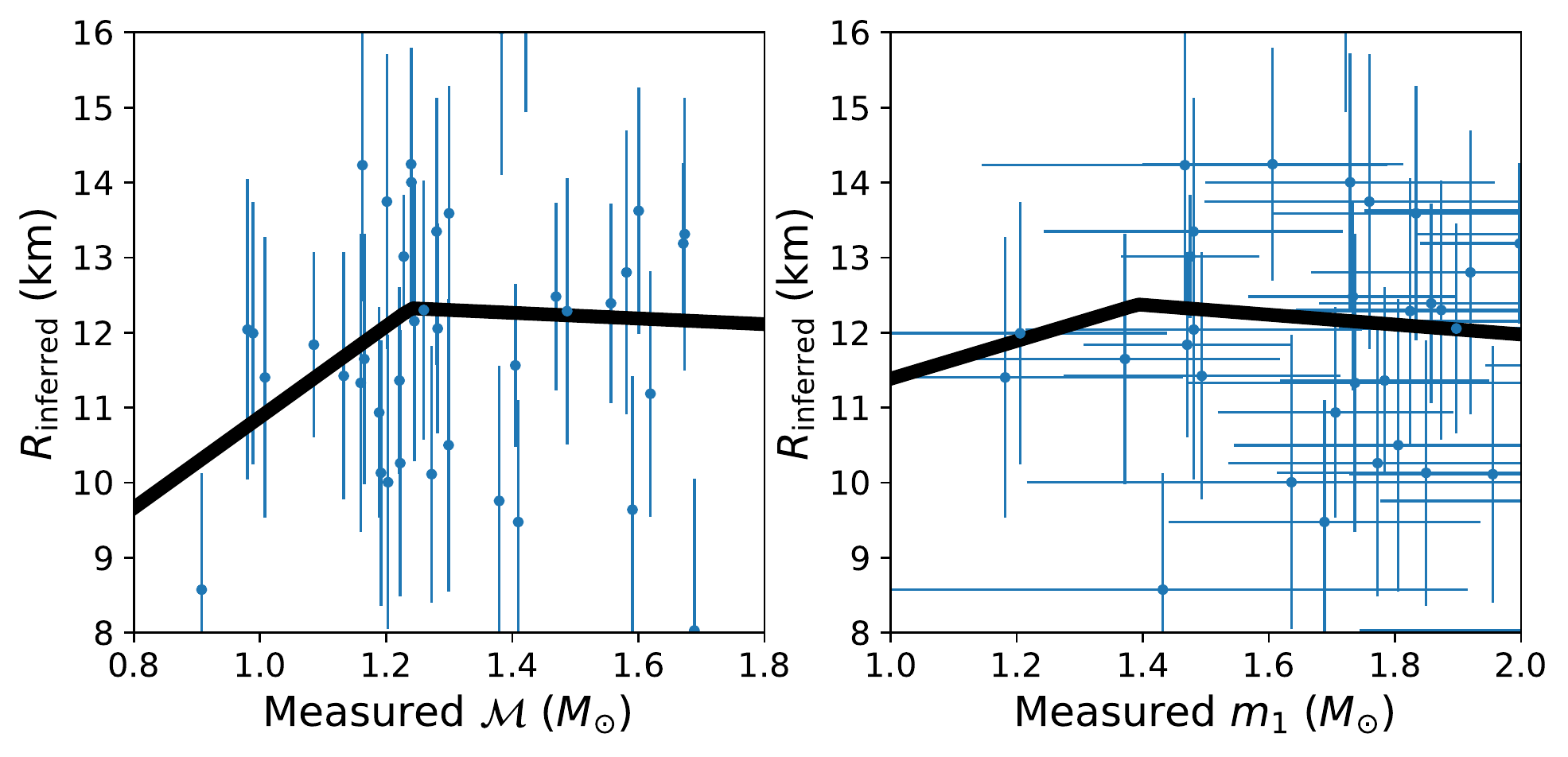}
	\caption{\label{fig:chirpmr} Left: an example of 40 simulated events in the $\mathcal M{-}R_{\rm inferred}$ plane for the same hybrid model shown in Fig.~\ref{fig:MChirpR}. Right: the same thing but in the $m_1{-}R_{\rm inferred}$ plane.  The solid lines in both plots are piecewise linear fits. }
\end{figure}

A simple scheme to search for kinks in the ${\mathcal M}{-}R_{\rm inferred}$ and $m_1{-}R_{\rm inferred}$ planes
is to fit the data to linear piecewise models $\mathcal R(\mathcal M|\mathcal P_0,\mathcal P_1,\mathcal P_2,\mathcal P_3)$ and $\mathcal R(m_1|P_0, P_1, P_2, P_3)$ respectively, with fit parameters $\mathcal P_i$ and $P_i$, and $\mathcal R$ given by Eq.~(\ref{eq:pw})~\footnote{The fitting was done with a 2-step $\chi^2$ minimization. First, we search for a minimum of $\chi^2$ on a course grid of parameters $\mathcal P_i$ or $P_i$ ([Lower bound, Upper bound, Number of grids in between] for the four parameters, respectively): $[10{\rm km},13{\rm km},6]$, $[1.0M_{\odot},2.0M_{\odot},10]$, $[-3{\rm km}/M_{\odot},3{\rm km}/M_{\odot},12]$, $[-45{\rm km}/M_{\odot},12{\rm km}/M_{\odot},50]$. We then use \texttt{scipy.optimize.leastsq} function in \texttt{python} to find the best fit parameters.}.
As an example, in Fig.~\ref{fig:chirpmr} we show one realization 
of 40 detections in the measured $\mathcal M{-} R_{\rm inferred}$ plane (left) and the measured 
$m_1{-} R_{\rm inferred}$ plane (right) for the hybrid model shown 
in Fig.~\ref{fig:MChirpR}. Also shown are the associated piecewise fits.  
We find that after repeating our simulations 500 times with 40 detections in each simulation, the majority of the simulations with HNSs return negative slope $\mathcal P_{3}$.  This is in qualitative agreement with the shape of the
allowed region in the $\mathcal M{-} R_{\rm inferred}$ plane shown in 
Fig.~\ref{fig:MChirpR} for hybrid data.  

\begin{figure}
	\centering
	\includegraphics[width=1.0\columnwidth]{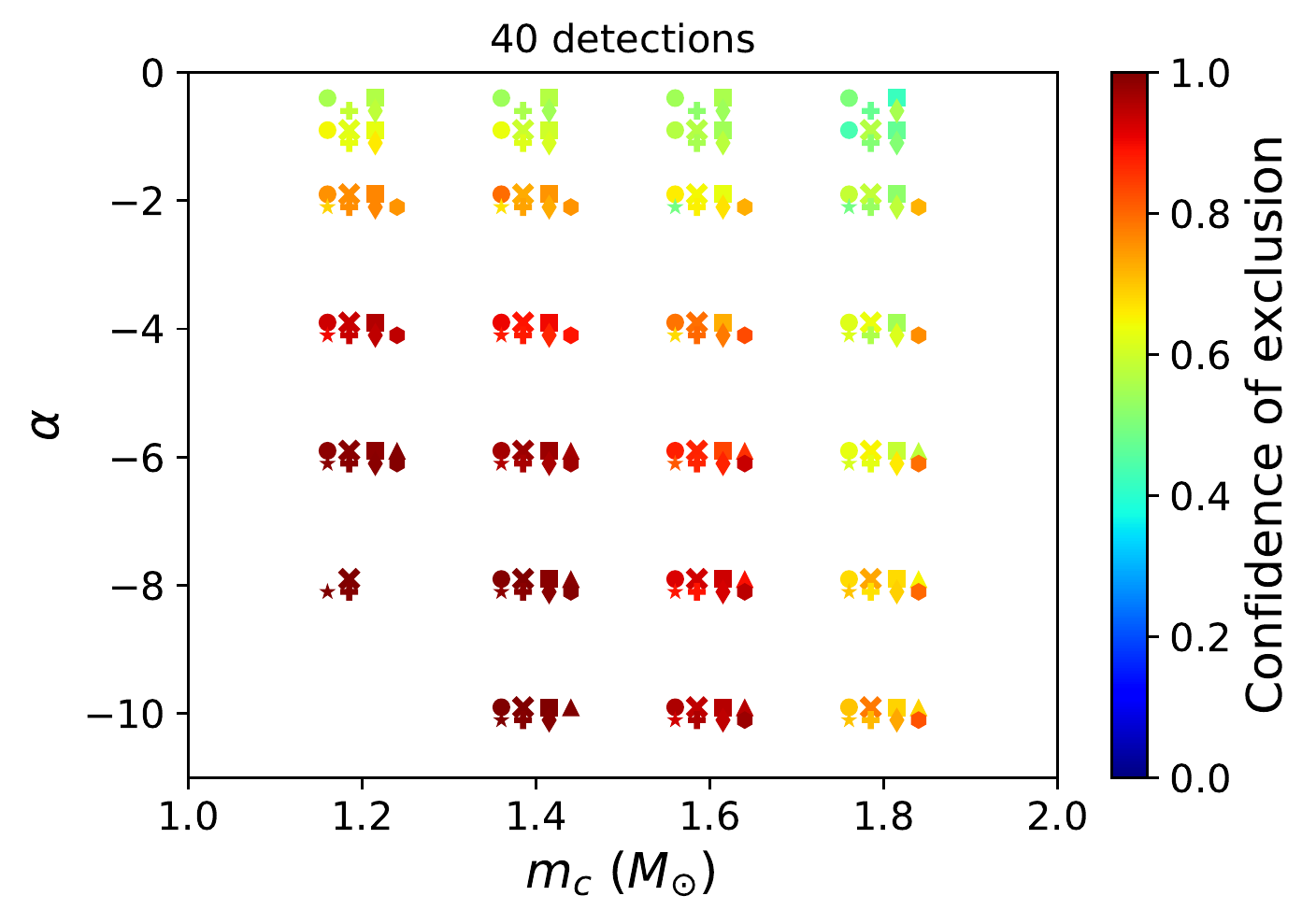}
	\caption{\label{fig:upperlimit}
		Confidence of exclusion for hybrid models as function of critical mass $m_c$
		and slope $\alpha$. The eight symbols represent our eight choices of parameters 
		$p_0$ and $p_2$, which are the same as those listed in Table.~\ref{eos}, with
		circle: model A, cross: model B, square: model C, triangle: model D, star: model E, plus: model F, diamond: model G, and hexagon: model H. 
	}
\end{figure}

In Figure~\ref{fig:upperlimit}, we show the fraction of 500 simulated {ensembles} which return $\mathcal P_{3}<0$ with hybrid injection data with slope $\alpha$ and critical mass $m_c$. {In each ensemble there are 40 simulated events.} 
This quantity measures the confidence a hybrid model with parameters $m_c$ and $\alpha$ can be excluded {if $\mathcal P_{3}>0$} with 40 BNS detections.  As is evident from the figure,
smaller critical masses $m_c$ and more negative slopes $\alpha$ can be excluded 
with greater confidence than larger $m_c$ and/or more positive $\alpha$. This is due to the fact that more negative $\alpha$ produce larger kinks in the allowed regions in the $\mathcal M{-} R_{\rm inferred}$ plane, making $\mathcal P_3$ more negative.  Likewise, larger $m_c$ leaves fewer events in the $[1,2] M_\odot$ range with masses $> m_c$, making the associated downward trend more difficult to resolve.  Nevertheless, it is noteworthy that already with 40 detections,
hybrid models with {$m_c \leq 1.6 M_\odot$} and $\alpha \leq -6 \ {\rm km}/M_{\odot}$
can be excluded at $>80\%$ confidence level if $\mathcal P_3 > 0$.

\begin{figure}
	\centering
	\includegraphics[width=1.0\columnwidth]{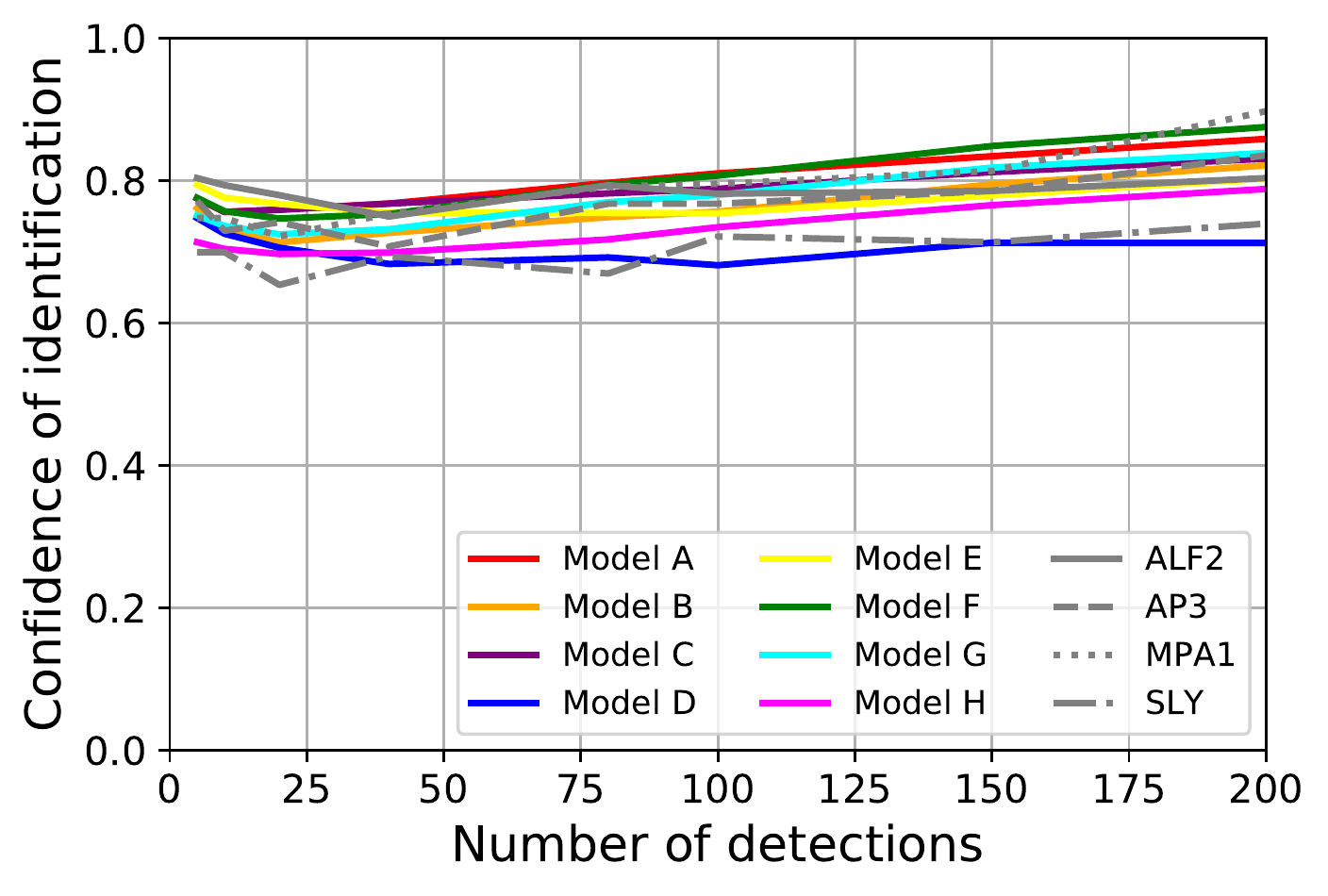}
	\caption{\label{fig:det}
		The confidence of identification of HNSs as a function of the number of detections
		with our eight different nuclear models as well as the ALF2, AP3, MPA1 and SLY EoS. 
	}
\end{figure}

On the other hand, with injection data from  nuclear models A-H, we find that the majority of 
our simulations return measured $\cal M$ slope $\mathcal P_3 > 0$ {or} measured $m_1$ slope discontinuity $P_2 - P_3 < 1 \ {\rm km}/M_\odot$. This selection criterion matches Fig.~\ref{fig:MChirpR}, where the nuclear model produced upwards trends in the $\mathcal M{-}R_{\rm inferred}$ plane as $\mathcal M$ increases and no kinks in the $m_1{-}R_{\rm inferred}$ plane.
In Fig.~\ref{fig:det} we plot unity minus the fraction of simulated nuclear {ensembles} which return {$\mathcal P_3 < 0$ and $P_2 - P_3 > 1 \ {\rm km}/M_\odot$} as a function of 
the number of {events} in each ensemble. 
This quantity can be 
interpreted as the confidence level of identifying a HNS. Fig.~\ref{fig:det} shows the confidence of identifying HNS as a function of the number of detections. Also shown in the figure is the confidence of identification for ALF2, AP3, MPA1 and SLY equations of state, which agrees with that obtained from their piecewise linear approximates.
Note the confidence of identification does not increase much as the number of events increases (and in fact even decreases slightly). This is due to our strict selection criteria, chosen to prevent misidentification of HNS,
as well as the fact that our piecewise linear fits only qualitatively describes the
distributions shown in Fig.~\ref{fig:MChirpR}.  
Nevertheless, for the most pessimistic nuclear model, model D, we can construct over 70\% confidence of identification with 40 observations if the observations return $\mathcal P_{3} < 0$ and $P_{2}-P_{3}>1 \ {\rm km}/M_\odot$. 
Since we do not know the actual nuclear MR curve, the results from model D can be conservatively taken as the confidence of identification when our method is applied to real data.  

\begin{figure}
	\centering
	\includegraphics[trim= 0 0 0 0 0,clip,scale=0.44]{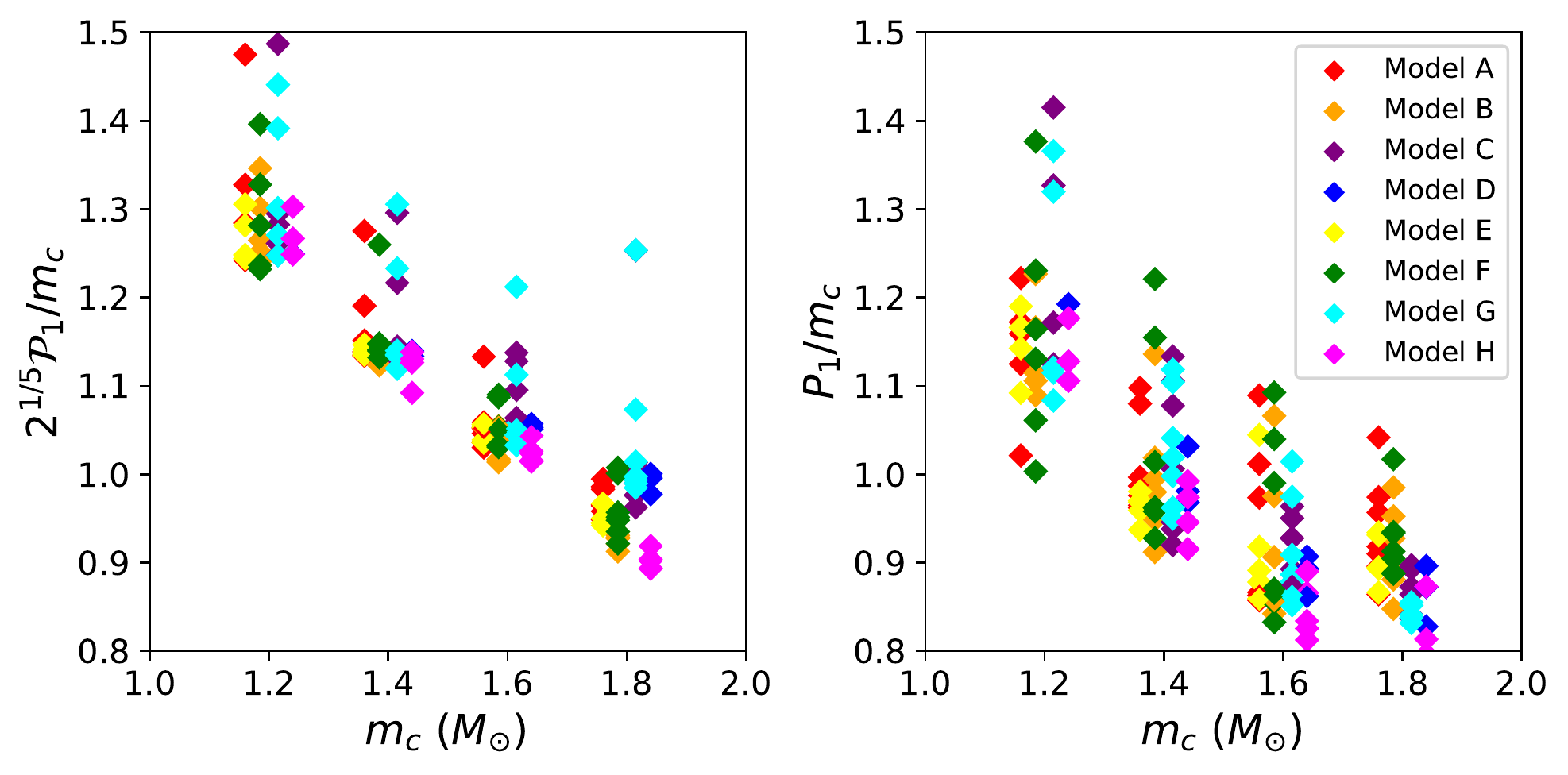}
	\caption{\label{fig:p1m1}
		Estimate of critical mass normalized to the injected value, $2^{1/5}\mathcal P_1/m_c$ (left) and $P_1/m_c$ (right), as a function of injected critical mass $m_c$, from 200 detections identified as having HNSs with our selection criterion.  
	}
\end{figure}

We conclude our analysis by discussing the reconstruction of the critical mass $m_c$ from our simulated data. 
From Fig.~\ref{fig:MChirpR} we see that kinks in the allowed regions in the $\mathcal M{-} R_{\rm inferred}$ and $m_1{-}R_{\rm inferred}$ plane occur at $\mathcal M = m_c/2^{1/5}$ and $m_1 = m_c$, respectively.  Therefore, for simulations that are identified as having HNSs,  $2^{1/5} \mathcal P_1$ and $P_1$ provide a rough estimate of $m_c$.
In Fig.~\ref{fig:p1m1}, we show  $2^{1/5} \mathcal P_{1}/m_c$ and $P_{1}/m_c$ as a function of the injected critical mass $m_c$, averaged over 500 simulations, each with 200 mergers.
The standard deviation of $\mathcal P_{1}$ is $0.05-0.7 M_\odot$ while the standard deviation of $P_{1}$ is $0.2-0.6 M_\odot$. 
In addition to the large statistical uncertainty, the figure also shows an estimate of the systematic errors. 
While the statistical uncertainty decreases over the number of detections, the systematic errors remain. Again, this is because our piecewise linear fits only qualitatively describes the distributions shown in Fig.~\ref{fig:MChirpR}.

\section{Discussion}While we have focused only on HNS with connected MR curves, we have also studied those disconnected MR curves.  In this case the 
allowed regions in the $\mathcal M{-} R_{\rm inferred}$ and $m_1{-}R_{\rm inferred}$ planes
have gaps, which reflect the presence of gaps in the associated MR curves.
Our present analysis cannot resolve these gaps, but can resolve the change in slope associated with the formation of a core with similar fidelity as reported above for connected models.
In the future, it would also be interesting to study the models with a mixed phase of nuclear and quark-matter. 
This scenario arises when the nuclear/quark matter surface tension is small \citep{Alford:2001zr}.
The presence of a mixed phase should soften kinks in MR curves, and correspondingly those in the $m_1{-}R_{\rm inferred}$ and $\mathcal M{-}R_{\rm inferred}$ planes. It remains to be seen 
how soft of a kink our analysis can detect.  

We note that the selection criterion used to identify HNSs is not unique. While using a stronger selection criterion might, for a given number of events, identify HNSs with greater confidence, it might also misidentify more HNSs as nuclear than a weaker criterion.

There is plenty of room to improve our data analysis.  Firstly, we assumed a uniform NS mass distribution between $[1,2]M_{\odot}$. Employing a different mass distribution will affect the fidelity in which
HNS can be probed. For example, if $m_c = 1.6 M_{\odot}$ and the NS mass distribution is narrowly distributed about $1.4 M_{\odot}$, identifying EM cores will be very difficult due to few high mass events.  As more and more gravitational wave events are detected, the mass distribution
should be better understood, and our analysis can be adjusted accordingly.

Additionally, our analysis of data in the $\mathcal M{-} R_{\rm inferred}$ and $m_1{-}R_{\rm inferred}$ planes can likely be improved.  For example, instead of fitted 
data to piecewise linear \textit{curves} $\mathcal R(\mathcal M|\mathcal P_0,\mathcal P_1,\mathcal P_2,\mathcal P_3)$ and $\mathcal R(m_1|P_0,P_1,P_2,P_3)$, it may be fruitful to fit instead to the \textit{density} of expected events,
which is determined by the mass distribution and the MR curve. This requires knowledge of the NS mass distribution.  Fitting to the density of events can likely 
ameliorate the aforementioned systematic errors on the measurement of the critical mass.

\begin{acknowledgments}We acknowledge valuable discussions with Katerina Chatziioannou, Ian Harry, Carl-Johan Haster, Philippe Landry, Jocelyn Read, and Salvatore Vitale. This work was supported by the Black Hole Initiative at Harvard University, which is funded by grants the John Templeton Foundation and the Gordon and Betty Moore Foundation to Harvard University.
\end{acknowledgments}

\bibliography{references2}

\end{document}